# Multimodal Stock Price Prediction

Furkan Karadaş[1], Bahaeddin Eravcı[2] and Ahmet Murat Özbayoğlu[2]
*[1]Department of Computer Engineering, TOBB University of Economics and Technology, Ankara, Türkiye*
*[2]Department of Artificial Intelligence Engineering, TOBB University of Economics and Technology, Ankara, Türkiye*
{fkaradas, beravci, mozbayoglu}@etu.edu.tr

Keywords: Financial Forecasting, Stock Market, Deep Learning, Deep Neural Networks, Finbert, Chatgpt.

Abstract: In an era where financial markets are heavily influenced by many static and dynamic factors, it has become increasingly critical to carefully integrate diverse data sources with machine learning for accurate stock price prediction. This paper explores a multimodal machine learning approach for stock price prediction by combining data from diverse sources, including traditional financial metrics, tweets, and news articles. We capture real-time market dynamics and investor mood through sentiment analysis on these textual data using both ChatGPT-4o and FinBERT models. We look at how these integrated data streams augment predictions made with a standard Long Short-Term Memory (LSTM model) to illustrate the extent of performance gains. Our study's results indicate that incorporating the mentioned data sources considerably increases the forecast effectiveness of the reference model by up to 5%. We also provide insights into the individual and combined predictive capacities of these modalities, highlighting the substantial impact of incorporating sentiment analysis from tweets and news articles. This research offers a systematic and effective framework for applying multimodal data analytics techniques in financial time series forecasting that provides a new view for investors to leverage data for decision-making.

## 1 INTRODUCTION

In the modern world of finance today, investors and fund managers find themselves confronting considerable challenges in making the most appropriate investment decisions possible in complicated and dynamic environments. From immediate-impact global economic events to political development market reflections and transformative industrial changes brought about by technological advancement, there have been so many continuous influences on financial markets that make predicting investments all the more difficult. Because of these reasons, traditional analysis methods often fail, and investors latch on to promising tools and frameworks that offer accurate and reliable forecasting.

Das et al. (Das, Behera, & Rath, 2018) and Peng et al. (Peng & Jiang, 2015) have focused on individual tweet content to predict stock prices using social media sentiment. This work is outstanding in terms of including broader data sources, news articles, and tweet sentiment data. However, unlike these studies, we enrich our sentiment analysis by incorporating engagement metrics such as tweet likes, retweets, comments and the tweeter's follower count into our feature extraction process. These processes provide a much more detailed view of market sentiment.

This study focuses on multimodal stock price prediction by integrating traditional financial metrics with data acquired from tweets from *Twitter (now X.com)* and news articles from *The New York Times*. We aim to capture real-time market dynamics and gauge investor sentiment by analyzing these data sources. Tweets provide non-moderated, real-time insights from the public, while news articles from *The New York Times* offer moderated, expert information. To measure market sentiment and excitement, we conducted sentiment analysis using both *ChatGPT-4o* and *FinBERT* to include (Araci, 2019). ChatGPT-4o is a recent, large-scale language model, whereas FinBERT is a specialized, smaller model tailored for financial text analysis. Our study evaluates the predictive models on three stocks representing various market states, including bull, bear, and neutral markets. We compare results from these integrated data sources with predictions made by a standard Long-Short-Term Memory (LSTM) model utilizing only price data as input. Then, we highlight the performance enhancements achieved through data diversity and advanced sentiment analysis.

Experimental results show that integrating social media data, such as tweets and news articles, with technical indicators significantly enhances the model's predictive accuracy, outperforming price data alone. Additionally, sentiment analysis using FinBERT and ChatGPT-4o produced very similar results, further validating the robustness of sentiment integration.

The following sections will provide a detailed methodology, including the experimental setup and analysis of results. We will also discuss the potential implications of multimodal data processing in financial predictions.

## 2 RELATED WORKS

Deep learning-based machine learning methods are successfully applied in various fields, such as medicine, computer vision, and telecommunications (Ahmedt-Aristizabal, Armin, Denman, Fookes, & Petersson, 2021). Due to their success in these areas, machine learning and deep learning have recently become popular among the methods used in finance, especially in financial forecasting. The application of machine learning and artificial intelligence techniques to financial data analysis and statistical analyses of temporal data (Boginski, Butenko, & Pardalos, 2005; Keskin, Yilmaz, & Ozbayoglu, 2021; Sezer, Gudelek, & Ozbayoglu, 2020; Tsay, 2005) provides investors with insights and recommendations when forming their portfolios.

LSTM (Hochreiter, 1997) is a variant of RNN capable of retaining short-term and long-term information. Deep learning researchers frequently choose LSTM networks for sequence learning. These models are mainly applied to time-series data and are employed across various domains, including Natural Language Processing (NLP), language modelling, translation, speech recognition, sentiment analysis, predictive analytics, and financial time series analysis (Gao, Chai, & Liu, 2017; Greff, Srivastava, Koutník, Steunebrink, & Schmidhuber, 2016; Wu, 2016).

A review of studies in the literature highlights the complexity and dynamism of financial markets, emphasizing that analyses based on a single feature can be misleading (Saha, Gao, & Gerlach, 2022). Pearson correlation reveals linear relationships but overlooks the diversity of financial data across different times. A single feature usually gives only a particular indication, while the interaction of several factors shapes financial markets. For example, focusing solely on stock prices can lead to paying attention to significant aspects such as the company's financial condition, management quality, industry trends, economic conditions, and competition.

Wang et al. (Wang, Yu, & Shen, 2020) utilize online financial reviews to determine the daily sentiment for each stock, finding a strong correlation between positive sentiment and an increase in the closing stock price. Akita et al. (Akita, Yoshihara, Matsubara, & Uehara, 2016) present a method for predicting stock prices using financial data metrics and text-based information. It introduces a strategy for forecasting stock prices that involves using distributed representations of news articles and examining the relationships between various companies operating within the same industry. Lavrenko et al. (Lavrenko et al., 2000) integrated stock price trends with financial news articles to predict market directions based on news content before these trends materialized. Another study analysed newspaper articles' sentiment added to the dataset and concluded that incorporating a sentiment-measuring feature improved model performance for the testing dataset (Forecast, 2021).

Besides news articles, social media is used in an array of studies (Cam, Cam, Demirel, & Ahmed, 2024; Das et al., 2018; Peng & Jiang, 2015). These implemented sentiment analyses on crawled tweets from Twitter along with stock data for forecasting in the stock market and concatenated with price data. While social media data, specifically tweets, has been incorporated into stock market forecasting models, these studies have typically limited their sentiment analysis to the textual content of tweets themselves.

Recent studies (Avramelou, Nousi, Passalis, & Tefas, 2024; Farimani, Jahan, & Fard, 2024; Taylor & Ng, 2024) examined multimodal deep learning for predicting financial markets, each using distinct methods and data types. Avramelou et al. (Avramelou et al., 2024) present a novel multimodal approach for deep reinforcement learning in financial trading. This specifically addresses the challenge of effectively combining diverse online data sources like news articles and social media websites. Their approach leverages embeddings to merge price and sentiment data, allowing the model to discover the best combinations of these elements for enhanced trading decisions. Taylor et al. (Taylor & Ng, 2024) explore a multimodal approach to stock price prediction, integrating news headlines and article sources with stock price percentage change data. Their study mainly examines how percentage change compares to raw price values in effectiveness, while also exploring how different combinations of these data types contribute to prediction accuracy. Farimani et al. (Farimani et al., 2024) propose an adaptive multimodal learning model for market price prediction, leveraging diverse data modalities to address the financial time series. The data sources for their models fetched from news content, sentiment from specialized newsgroups, and technical indicators.

## 3 METHOD

### 3.1 Dataset

The dataset for this study comprises three selected stocks: Walmart Inc., Walt Disney Co., and Microsoft. Despite analyzing a few stocks, we ensured a more representative dataset by selecting companies with significantly different market capitalizations, sectors, and behaviors in market states. *Table 1* lists each stock's sector information, market state, and capitalization.

Table 1: Selected Stocks

| Stock Name | Sector | State | Market Cap 2024 |
|---|---|---|---|
| Walmart Inc. | Consume Defensive | Neutral | 643.86B $ |
| Walt Disney Co. | Communication Services | Bear | 170.71B $ |
| Microsoft | Technology | Bull | 3.09T $ |

With their different market conditions, these stocks represent bull, bear, and neutral states. Bull market stocks include rising prices and investors' confidence while falling prices and economic challenges characterize bear market stocks. Neutral market stocks have a stable nature with a moderate fluctuation in price to give a balanced perspective. These stocks represent various market dynamics and how those dynamics may play into investment strategy. We collected comprehensive financial data for each company from Yahoo Finance, including historical price movements and other relevant financial indicators. The data gathered for each stock ranges from 2018 to 2023. The parameters chosen for analysis are Date, Open, High, Low, Close, Adjusted Close, and Volume.

We collected tweets relevant to each stock to understand public opinion on market dynamics by filtering for specific keywords. Keywords included the company name or stock ticker symbol, such as "Microsoft or MSFT." We required each tweet to have at least 100 likes to prioritize tweets with higher engagement. *Table 2* shows the total number of tweets for each stock. These tweets and the stock price data were gathered from 2018 to 2024.

Table 2: Number of Tweets for Stocks

| Stock | Number of Tweets |
|---|---|
| Walmart Inc. | 31.555 |
| Walt Disney Co. | 183.406 |
| Microsoft | 53.446 |

We included another data source for analyzing market insights with expert and moderated opinions to augment our market understanding. We collected news articles from The New York Times using its API to access news articles about each stock's significant events and developments that may affect market sentiment or stock performance. We filtered the news articles using the exact keywords used for the tweets. *Table 3* presents the total number of news articles for each stock collected over the same period as the tweet data.

Table 3: Number of News Articles for Stocks

| Stock | Number of Tweets |
|---|---|
| Walmart Inc. | 2.930 |
| Walt Disney Co. | 5.821 |
| Microsoft | 3.629 |

### 3.2 Preprocessing Data

In the preprocessing data phase, due to the non-uniform range of values in the historical trading data, Min-Max normalization will be applied to scale the data to a range between 0 and 1 before inputting it into the LSTM model.

We cleaned the data using various techniques to ensure the quality and relevance of text within tweets and news articles. This step removed unnecessary content or text around URLs, hashtags, mentions, reserved words, emojis, and smileys. Additionally, we eliminated stop words, punctuation, special characters, and numbers, as they do not contribute meaningful information to the text. Further, we removed any extra spaces while converting all text to lowercase. These steps were essential for the dataset, making it suitable for the subsequent analysis.

To address missing data, we applied a data-filling method to ensure the continuity and completeness of the dataset. Using forward-filling techniques, we effectively imputed missing values while preserving the dataset's integrity.

## 3.3 Feature Extraction

Firstly, we selected two models to analyse the sentiment of the tweets and news: FinBERT (Araci, 2019) and ChatGPT-4o. These two models bring significant innovations to natural language processing (NLP). FinBERT is an adaptation of the BERT architecture and is specifically trained to process financial texts, which means it allows better and quicker insights from financial documents like market analyses. The other model is ChatGPT-4o, which is a general language model used to answer many different kinds of questions effectively. Besides FinBERT, it is trained not only on financial data but in more and larger open-source data for human interaction; hence, it will generate creative solutions during conversations. Each tweet was input into these models to assess its sentiment, assigning a score ranging from -1 to 1. A score of -1 represents a highly negative sentiment, 0 signifies a neutral sentiment, and 1 denotes a highly positive sentiment. Additionally, the models provided an accuracy percentage for the sentiment evaluation, allowing us to quantify the models' confidence in classifying sentiment accurately.

Furthermore, we introduced a weighted sentiment score to enhance our sentiment analysis. This method considers not only the sentiment score of each tweet but also additional engagement metrics: likes, retweets, and comments for tweets, along with the follower count of the user who posted the tweet. We defined the tweet interaction ($T_i$) such that

$$T_i = \alpha * T_r + \beta * T_l + \gamma * T_c \quad (1)$$

where $T_r$ number of retweets for the tweet, $T_l$, number of likes for the tweet, $T_c$, and number of comments for the tweet and also $\alpha, \beta, \gamma$ are hyperparameters reflecting the weights assigned to each metric. In this study, we set these parameters to 0.3 as an initial estimate without any optimization.

To account for the impact of the user who posted the tweet, we calculated the user influence ($U_i$) based on their follower count, capturing the potential reach of their messages. This is expressed as

$$U_i = \delta * F_C \quad (2)$$

where $F_C$ number of followers for the user who posted the tweet, $\delta$ is a hyperparameter determining the influence of follower counts, chosen as 0.1 in this study.

The sentiment ($S$) is derived by multiplying the sentiment label ($S_l$) with the accuracy percentage of the sentiment classification ($S_c$), ensuring that the calculated sentiment reflects both its evaluated value and the confidence level.

$$S = S_l * S_c \quad (3)$$

Total tweet interaction ($TT_i$) aggregates the total engagement across retweets, likes, and comments to assess each tweet's overall impact

$$TT_i = T_r + T_l + T_c \quad (4)$$

where $T_r$ number of retweets for the tweet, $T_l$, number of likes for the tweet, $T_c$, number of comments for the tweet.

By employing these comprehensive formulas, we could reflect not just the content of the tweets but also their potential influence and effectiveness within the social media landscape. The weighted sentiment ($WS$) formula integrates all these factors and is calculated as follows:

$$WS = \frac{T_i * U_i * S}{TT_i} \quad (5)$$

Our approach incorporates weighted sentiment analysis, which augments tweet sentiments by considering tweet engagement (retweets, likes, comments) and user influence (follower count). This is represented by the hyperparameters α, β, γ, and δ which are tuned to optimize model performance and provide a more nuanced assessment of market sentiment, as high engagement and influential users are weighted more heavily. This approach is expected to more accurately reflect the true impact of the influence and engagement of tweets on stock prices compared to just tweet content's sentiment.

To enhance our dataset, we incorporated financial technical indicators and sentiment analysis. These features include the Relative Strength Index (RSI) and the Simple Moving Average (SMA), both of popular in financial analysis. The RSI measures the speed and change of price movements to determine whether a stock is overbought or oversold, while the SMA smooths price data to help identify trends over specified time periods.

To integrate the news articles, tweets, and price data, we faced the challenge of aligning datasets that operate on different timeframes. Given that the price data is recorded daily, while there can be hundreds of tweets and news articles within a single day, we needed to synchronize these varying data frequencies. We calculated the average sentiment label and accuracy percentage for all tweets and news articles generated daily to achieve this. Additionally, we included the number of tweets and news articles for the respective day in the dataset. This approach allowed us to convert all data into a daily format, ensuring that each stock's sentiment analysis corresponds accurately with the price movements and providing a cohesive dataset for our analysis.

### 3.4 Model

Our predictive model uses a standard Long-Short-Term Memory (LSTM) architecture, chosen specifically for its general applicability in handling sequential data and its robustness in time series forecasting. We focused on predicting stock closing prices, which are crucial indicators in financial decision-making.

The LSTM model consists of a single layer with 256 units. We employed the ReLU activation function and the Adam optimizer with a learning rate 0.001. To optimize computational resources and facilitate effective learning, we selected a batch size of 128. The model was trained over 100 epochs.

## 4 PERFORMANCE EVALUATION AND DISCUSSIONS

### 4.1 Training Strategy

Our model was trained using data from 2018 to 2022, while data from 2023 was reserved for testing. This temporal split allows the model to learn from a substantial period of historical data before being evaluated on more recent data patterns and unseen market dynamics. We selected the Mean Squared Error (MSE) for the loss function. To enhance the reliability of our findings, we trained the model 10 times for both the training and testing phases and then averaged the results. This iterative approach minimizes the effects of any random variances or anomalies in the data, ensuring that the performance metrics reflect a more stable and generalized model performance.

### 4.2 Computational Model Performance

In evaluating the performance of our model, designed to predict stock closing prices, we utilized two key metrics: R-squared (R2) and Mean Absolute Error (MAE). A separate model was trained and tested independently for each stock to account for different equities' unique characteristics and behaviours. Both metrics were calculated using the predicted and actual closing prices.

*Table 4* presents the performance of stock price prediction models using FinBERT sentiment analysis, evaluated across various feature combinations.

For **Walmart**, the combination of price data with RSI and SMA yielded the best results ($R^2$ = 0.9282, MAE = 0.0181). Similar observations were made for Microsoft. The highest performance was achieved using price data combined with RSI and SMA ($R^2$ = 0.9644, MAE = 0.0196).

Table 4: Model Performance $R^2$ and MAE Scores (FinBERT Sentiment)

| Metrics / Features | *Walmart Inc.* | | *Walt Disney Co.* | | *Microsoft* | |
|---|---|---|---|---|---|---|
| | *$R^2$* | *MAE* | *$R^2$* | *MAE* | *$R^2$* | *MAE* |
| Prices (Baseline) | 0.9282 | 0.0181 | 0.8915 | 0.0166 | 0.9497 | 0.0233 |
| Prices-RSI-SMA | **0.9357** | 0.0166 | 0.8899 | 0.0167 | **0.9644** | 0.0196 |
| Prices-News | 0.8893 | 0.0229 | 0.8605 | 0.0192 | 0.8118 | 0.0355 |
| Prices-News-RSI-SMA | 0.9041 | 0.0212 | 0.8291 | 0.0214 | **0.9604** | 0.0206 |
| Prices-Tweets | 0.8913 | 0.0233 | 0.8888 | 0.0168 | 0.9494 | 0.0232 |
| Prices-Tweets-RSI-SMA | 0.9179 | 0.0194 | 0.8497 | 0.0198 | **0.9603** | 0.0206 |
| Prices-Tweets-News | 0.8893 | 0.0229 | 0.8285 | 0.0216 | 0.9396 | 0.0250 |
| Prices-Tweets-News-RSI-SMA | 0.9041 | 0.0212 | 0.8454 | 0.0200 | **0.9640** | 0.0193 |
| Prices-Weighted-Tweets | 0.8724 | 0.0258 | **0.9086** | 0.0150 | **0.9516** | 0.0228 |
| Prices-Weighted-Tweets-RSI-SMA | 0.8841 | 0.0241 | 0.7978 | 0.0228 | **0.9609** | 0.0204 |
| Prices-Weighted-Tweets-News | 0.8731 | 0.0248 | 0.8709 | 0.0183 | 0.9433 | 0.0246 |
| Prices-Weighted-Tweets-News-RSI-SMA | 0.9099 | 0.0205 | 0.8626 | 0.0189 | 0.9456 | 0.0242 |

In contrast, incorporating sentiment features improved model performance for **Walt Disney Co.** predictions. For instance, the model that combined price data with the weighted sentiment score of tweets achieved higher R² scores (R² = 0.9086, MAE = 0.0150) than just using price data (R² = 0.8915, MAE = 0.0166), suggesting that sentiment data captures valuable information about market perceptions and investor sentiment specific to Walt Disney.

It should be noticed that for **Microsoft**, the model with combined data achieved better results with tweets, news articles, and technical indicators (R² = 0.9640, MAE = 0.0193) compared to using only price data (R² = 0.9497, MAE = 0.0233). This highlights the significance of sentiment analysis for Microsoft's stock predictions. Sentiment data can be beneficial when combined with price data, enhancing the model's ability to capture market trends more effectively.

*Table 5* summarizes the performance of models utilizing ChatGPT-4o sentiment analysis features across different stocks, revealing notable variations depending on the features used.

For **Walmart**, the best results were again achieved with technical indicators like RSI and SMA (R² = 0.9423, MAE = 0.0155). Despite this being the optimal outcome, when tweet sentiment data and technical indicators are added, it yields better results (R² = 0.9352, MAE = 0.0167). than using only price data (R² = 0.8918, MAE = 0.0226). For **Walt Disney**, it was observed that using only price data was more effective than including other features, such as sentiment features from tweets, for improving the overall model performance. The model performed best for **Microsoft** using traditional technical indicators with tweet sentiment data (R² = 0.9588, MAE = 0.0211).

Both demonstrated strong performance when comparing the FinBERT and ChatGPT-4o sentiment analysis features. These indicators consistently led to predictive performance across all stocks in both models. Adding sentiment data, like news and tweets, contributed positively in both models, though the improvements were modest. For stocks like Walt Disney, sentiment features provided slight enhancements in the score, but the overall impact of sentiment data was similar in both models. FinBERT and ChatGPT-4o performed well, with technical indicators playing the dominant role and sentiment features adding subtle yet consistent value.

Table 5: Model Performance R² and MAE Scores (ChatGPT Sentiment)

| Metrics / Features | Walmart Inc. R² | Walmart Inc. MAE | Walt Disney Co. R² | Walt Disney Co. MAE | Microsoft R² | Microsoft MAE |
|---|---|---|---|---|---|---|
| Prices (Baseline) | 0.8918 | 0.0226 | **0.9157** | 0.0143 | 0.9466 | 0.0241 |
| Prices-RSI-SMA | **0.9423** | 0.0155 | 0.8754 | 0.0180 | **0.9529** | 0.0228 |
| Prices-News | 0.8763 | 0.0248 | 0.8778 | 0.0177 | 0.9319 | 0.0275 |
| Prices-News-RSI-SMA | **0.8930** | 0.0226 | 0.7492 | 0.0268 | **0.9547** | 0.0221 |
| Prices-Tweets | **0.8934** | 0.0231 | 0.9026 | 0.0157 | **0.9525** | 0.0225 |
| Prices-Tweets-RSI-SMA | **0.9352** | 0.0167 | 0.8614 | 0.0192 | **0.9588** | 0.0211 |
| Prices-Tweets-News | 0.8759 | 0.0248 | 0.8362 | 0.0209 | 0.9351 | 0.0258 |
| Prices-Tweets-News-RSI-SMA | **0.8994** | 0.0221 | 0.7385 | 0.0273 | **0.9555** | 0.0218 |
| Prices-Weighted-Tweets | **0.9212** | 0.0196 | 0.8984 | 0.0160 | **0.9510** | 0.0230 |
| Prices-Weighted-Tweets-RSI-SMA | **0.9174** | 0.0194 | 0.8734 | 0.0181 | **0.9467** | 0.0243 |
| Prices-Weighted-Tweets-News | 0.8809 | 0.0236 | 0.7992 | 0.0230 | **0.9481** | 0.0234 |
| Prices-Weighted-Tweets-News-RSI-SMA | **0.9086** | 0.0207 | 0.7149 | 0.0286 | **0.9505** | 0.0233 |

## 4.3 Market Simulation

We implemented real-world stock trading using the strategy outlined by Lavrenko et al. (Lavrenko et al., 2000), which is defined as follows

$$r_{c_n}(t) = \frac{closing^{pred}_{c_n}(t) - opening^{true}_{c_n}(t)}{opening^{true}_{c_n}(t)} \qquad (6)$$

$$gain_{c_n}(t) = \begin{cases} buy \to sell & (r_{c_n}(t) > 0) \\ sell \to buy & (r_{c_n}(t) < 0) \end{cases} \quad (7)$$

where buy → sell denotes a transaction purchasing stocks at the opening price, and sell → buy denotes a transaction selling at the opening price. Furthermore, shares are purchased at the closing price if the opening price decreases by 2% relative to the predicted closing price. In other cases, if a profit of 2% is achieved based on the price at which the stock was initially bought, shares are sold at either the opening or closing price, depending on which offers the realized gain.

The market simulation results in *Table 6* illustrate the percentage gains achieved in various feature combinations, starting with an initial capital of 1 million dollars for each stock. The table shows the results of the profit we made at the end of the year.

Prices and news articles features achieved the highest score, 12.3973%, for **Walmart** using FinBERT sentiment. For **Walt Disney**, we observe that the tweets obtained the highest score, 1.61629% when weighted sentiment analysis was used, which was done using FinBERT sentiment analysis. When we look at the **Microsoft** results, tweets, news articles, and technical indicators, we see that they reached a peak score of 42.1134% with sentiment analysis using FinBERT.

These results demonstrate that incorporating technical indicators and sentiment analysis leads to higher returns than price data. In particular, adding features such as RSI, SMA, and sentiment data (especially news articles and tweets) has been observed to improve model performance consistently. When technical indicators and sentiment analysis features are added to price data, the results show improved returns across all three stocks. While price data provides a strong baseline, combining technical indicators and sentiment analysis allows the models to capture market trends more effectively and generate higher profits. Thus, enhancing the models with these additional features proves to be a more effective strategy for maximizing gains in the simulation.

Table 6: Market Simulation Score

| Sentiment Model / Features | Walmart Inc. | | Walt Disney Co. | | Microsoft | |
|---|---|---|---|---|---|---|
| | *FinBERT* | *ChatGPT 4o* | *FinBERT* | *ChatGPT 4o* | *FinBERT* | *ChatGPT 4o* |
| Prices (Baseline) | 9.0134 | 10.464 | -0.6719 | **-0.8018** | 27.5833 | 24.8890 |
| Prices-RSI-SMA | 5.2395 | 6.1929 | -10.6415 | -11.3467 | **30.9224** | **26.3306** |
| Prices-News | **12.3973** | 7.5495 | -2.7224 | -3.5872 | 28.2539 | 15.2766 |
| Prices-News-RSI-SMA | 5.2205 | 6.0573 | -2.5248 | -6.0232 | 30.9294 | **28.4940** |
| Prices-Tweets | 7.3985 | 6.9832 | -2.9358 | -6.5354 | 23.8664 | **26.1302** |
| Prices-Tweets-RSI-SMA | 6.8397 | 7.9131 | -6.1595 | -8.1733 | 30.7342 | 27.8191 |
| Prices-Tweets-News | **10.6523** | 9.2494 | -2.1445 | -4.3265 | **35.5574** | 29.1749 |
| Prices-Tweets-News-RSI-SMA | 5.8697 | 6.9430 | -6.2386 | -6.2184 | **42.1134** | **29.6521** |
| Prices-Weighted-Tweets | 3.6977 | 3.8771 | -2.0255 | -3.3259 | 23.3933 | 21.2062 |
| Prices-Weighted-Tweets-RSI-SMA | 2.8009 | 7.7851 | -9.1950 | -8.7334 | **33.6941** | 22.6620 |
| Prices-Weighted-Tweets-News | **11.7886** | 10.501 | **1.6162** | -1.3866 | 27.7016 | 19.7598 |
| Prices-Weighted-Tweets-News-RSI-SMA | 5.7785 | 6.5935 | -3.5114 | -4.7983 | 25.5040 | **29.7701** |

## 5 CONCLUSIONS

This paper demonstrates the advantages of a multimodal approach to stock price prediction, effectively combining traditional financial metrics with sentiment analysis from tweets and news articles.

To evaluate the effectiveness of our approach, we conducted experiments using market simulation techniques. The results revealed that incorporating sentiment analysis of textual data (tweets and news articles) into the model improved its performance compared to using price data alone.

This paper has demonstrated how the use of traditional financial data in conjunction with multiple sources of text-based data, like tweets or news articles, will lead to more accurate financial forecasting. By using insights from sentiment analysis of text-based resources like tweets and news, this study also emphasizes how crucial it is to comprehend market sentiment and how it affects changes in stock prices.

In other words, a multimodal approach to financial data analysis can enhance prediction accuracy and result in more effective trading strategies.

To improve profit-making capacities, we want to include macroeconomic information in future work, such as GDP growth rates, inflation rates, and unemployment statistics.

## APPENDIX

We used a specific prompt to fetch the sentiment in tweets and news articles from ChatGPT-4o. Below is the prompt used to guide the ChatGPT-4o:

"You are an experienced financial analyst tasked with analyzing tweets and news related to a specific stock to gauge the overall sentiment and potential impact on the stock's price.

For each given tweet or news snippet about the target stock, please:

1. Carefully consider the sentiment expressed, looking at factors like:
    - Positive or negative language and tone
    - Mentions of financial performance, profits/losses, business developments
    - Discussion of stock price movements, investor confidence
    - Overall implications of the content for the stock
2. Based on your analysis, provide the sentiment label (positive, negative, or neutral) and a sentiment score (between 0 and 1) representing the probability of the sentiment label (e.g., a score of 0.8 for a negative label means there is an 80% probability that the tweet is negative).
3. Provide the sentiment score for each text item, along with a one-sentence explanation for your score.

Please look at the 'Content' column and analyze each row. Then, add columns for sentiment label and scoring (between 0 and 1) in the file.

Finally, please summarize your findings with:
- The average sentiment score across all the tweets/news
- A brief paragraph highlighting the key positive and negative drivers of sentiment based on the text provided. What are the main factors influencing the overall sentiment?

Remember to consider the financial and investing context carefully, not just generic sentiment. Focus on how the information may impact the stock and investor perceptions.

You should add the sentiment label and score in the current file for me."